\documentclass[twocolumn,floats,prl,aps,showpacs]{revtex4}
\usepackage{graphicx}
\usepackage{amsfonts}
\usepackage{amsmath}
\usepackage{amssymb}
\usepackage{exscale}
\usepackage{color}

\begin{document}

\title{Energy of $N$ Cooper pair by analytically solving Richardson-Gaudin equations}
\author{ Michel Crouzeix$^1$, Monique Combescot$^2$}
\affiliation{(1) Institut de Recherche Math\'ematiques de Rennes, Universit\'e de Rennes
1, Campus de Beaulieu, 35042 Rennes cedex, France}
\affiliation{(2) Institut des NanoSciences de Paris, Universit\'e Pierre et Marie Curie,
CNRS, Tour 22, 4 place Jussieu, 75005 Paris}
\date{\today }

\begin{abstract}
This Letter provides the solution to a yet unsolved basic problem of Solid State Physics:  the ground state energy of an arbitrary number of Cooper pairs interacting via the Bardeen-Cooper-Schrieffer potential. We here break a 50 year old math problem by analytically solving Richardson-Gaudin equations which give the exact energy of these $N$ pairs via $N$ parameters coupled through $N$ non-linear equations. Our result fully supports the standard BCS result obtained for a pair number equal to half the number of states feeling the potential. More importantly, it shows that the interaction part of the $N$-pair energy depends on $N$ as $N(N-1)$ only from $N=1$ to the dense regime, a result which evidences that Cooper pairs interact via Pauli blocking only. \end{abstract}

\author{}
\maketitle
\date{\today }

Superconductivity\,\cite{Schrieffer,Fetter,Tinkham,Leggett} is one of the most fascinating phenomena of Solid State Physics. Its  physical understanding stayed a major problem for half a century \cite{onnes}. The first step towards this understanding came from
H.~Fr\"{o}hlich\,\cite{Frol} who pointed out that two electrons with opposite spins can attract each other via the ion motion. A few years later, L.~Cooper showed\,\cite{Cooper} that, no matter how weak this attraction is, two electrons with opposite spins can form a bound state when added to a Fermi sea in an energy layer where a small attracting potential acts - we will call this layer "potential layer" in the following. To handle more than one pair is clearly difficult because this imposes to control Pauli blocking between paired fermions. A way to overcome this difficulty is to turn to the grand canonical ensemble, with a pair number not fixed, as proposed by Bardeen, Cooper and Schrieffer (BCS).
Using a wave function ansatz - based on the idea that electron pairs are bosonic particles, so that they are likely to condense all into the same state - they derived\,\cite{BCS}, through a minimization of the Hamiltonian mean-value, the condensation energy for electrons filling half the potential layer. 

A few years after BCS work, R.W.\,Richardson\,\cite{Ric1,Ric2,Ric3} and M.\,Gaudin\,\cite{Gaud1,Gaud2},
made an important step in showing that the BCS Hamiltonian leads to one of the very few exactly solvable Schr\"{o}dinger equations. Indeed, the \emph{exact} energy of $N$ up/down spin electrons paired by the so-called "reduced BCS potential", 
\begin{equation}
V_{BCS}=-V\sum w_{\textbf{p}'}w_{\textbf{p}}a_{\textbf{p}'\uparrow}^\dag a_{-\textbf{p}'\downarrow}^\dag a_{-\textbf{p}\downarrow}a_{\textbf{p}\uparrow}
\end{equation}
with $w_{\textbf{p}}=1$ for $\varepsilon_{\textbf{p}}$ in the potential layer $(\varepsilon_{F_0}, \varepsilon_{F_0}+\Omega)$, reads as a sum of $N$ complex quantities $R_j$ solution of $N$ algebraic equations. Although this is a significant advance compared to solving a second order differential equation for $N$-body wave function, the resolution of these $N$ equations still is a formidable math problem which, over the past half century, stayed unsolved for arbitrary $N$ and potential strength. Through an elegant electrostatic analogy\,\cite{Ric4}, Richardson succeeded to recover the BCS condensation energy in the large $N$ limit. For $N$ small enough, these equations are commonly approached numerically to understand the physics of superconductor granules\,\cite{gran1,gran2,gran3}.

Cooper pairs are composite bosons quite different from the semiconductor excitons we extensively studied over the last decade\,\cite{Monique}. Through our exciton studies, we however understood that the many-body physics of composite bosons is mainly driven by the Pauli exclusion principle between the particle fermionic components. In order to microscopically control the effect of Pauli 
blocking when the Cooper pair number increases, it is necessary to stay in the canonical ensemble with both, the number of pairs and the number of states available for pairing, fixed. This is why we decided to tackle  these Richardson-Gaudin equations again in order to solve them analytically for arbitrary  $N$.

To grasp the trend induced by Pauli blocking on the pair binding energy, we first considered two pairs: even this $N=2$ problem had no known solution although this definitely is the next problem to tackle after $N=1$ studied by Cooper. A year ago, we showed\,\cite{PCC} that the \emph{exact} energy for two pairs reads as 
\begin{equation}\label{e1}
E_2=2E_1+\frac{2}{\rho}\big(1+\frac{2\sigma}{1-\sigma}\big)\frac{\tan\theta/2}{\theta/2},
\end{equation}
with $\theta$ such that
\begin{equation}\label{e2}
\frac{2\theta\sin\theta}{1-2\sigma\cos\theta+\sigma^2}=\frac{2}{\rho\, \Omega\, \sigma}\equiv \gamma.
\end{equation}
$E_1=2\varepsilon_{F_{0}}-\varepsilon_c$ is the single pair energy found by Cooper. $\varepsilon_{F_{0}}$ is the Fermi energy of the electrons which do not feel the potential.
$2\sigma/(1\!-\!\sigma)=\varepsilon_c/\Omega$ is the single pair binding energy in unit of the potential extension $\Omega$. The dimensionless parameter $\sigma=\exp(-2/\rho V)$, with $\rho$ being the density of states
taken as constant in the potential layer, is sample volume free, $\rho$ and $1/V$ linearly increasing with volume.

This 2-pair energy gives hints to understand the effects of Pauli blocking on Cooper pairs. In a large sample, $\rho\to\infty$, so $\theta\to\ 0$.
Difference between the energies of two correlated pairs $E_2$ and two single pairs $2E_1$ has a naive contribution $2/\rho$ which comes from the fact that the second pair must occupy the up/down states $(\varepsilon_{F_{0}}+1/\rho)$ just above the $\varepsilon_{F_{0}}$ Fermi level. $E_2-2E_1$ also has a more subtle contribution $2\varepsilon_c/N_\Omega$ where $N_\Omega=\rho\Omega$ is the number of states in the potential layer from which paired electrons are formed. This $2\varepsilon_c/N_\Omega$ contribution brings the 2-pair correlation energy from the two single-pair value $2\varepsilon_c$ down to $2\varepsilon_c(1\!-\!1/N_{\Omega})$. This binding energy decrease is induced by the \emph{moth-eaten effect} coming from the Pauli exclusion principle between composite bosons\,\cite{Monique}. It
is better understood by writting $\varepsilon_c$ as $N_{\Omega}\varepsilon^*$ where $\varepsilon^*=2\sigma/(1\!-\!\sigma)\rho$ is the contribution of each of the $N_\Omega$ empty pair states in the potential layer. 
When a second pair is added, the number of states available for pairing decreases from $N_\Omega$ to $N_\Omega\!-\!1$; so the binding energy must decreases, due to Pauli blocking, from $N_\Omega \varepsilon^*$ to $(N_\Omega\!-\!1)\varepsilon^*$, as we find.

The purpose of this Letter is to see how this microscopic understanding of the Cooper pair correlation energy extends to $N>2$. We here present an analytical solution of  Richardson-Gaudin equations for the energy of an arbitrary number $N$ of up/down spin electrons paired by the BCS potential. We find that the $N$-pair energy takes a remarkably compact form
\begin{equation}\label{e3}
E_N= NE_1+\frac{N(N-1)}{\rho}\frac{1+\sigma}{1-\sigma}
\end{equation}
within terms in $(N/{\rho})^n$ free from sample volume and thus negligible in the thermodynamic limit, in front of the two volume linear terms of $E_N$. 

This result reduces not only to Eq.\,\eqref{e1} for $N=2$ but also to the energy obtained by Bardeen, Cooper and Schrieffer for $N_{BCS}=N_\Omega/2$ which corresponds to a potential extending symmetrically on both sides of the normal electron Fermi sea. Indeed, Eq.\,\eqref{e3} gives the condensation energy as 
\begin{equation}\label{e3'}
E_{N}^{(0)}-E_N= N\varepsilon_c\left[1-\frac{N-1}{N_\Omega}\right],
\end{equation}
where $E_{N}^{(0)}=2\left[N\varepsilon_{F_{0}}+N(N-1)/2\rho\right]$ is the energy of $N$ "normal" electrons added above  $\varepsilon_{F_{0}}$, for a constant density of states $\rho$. This condensation energy reduces to $\varepsilon_cN_{BCS}/2$ for half filling which is the BCS value, $\rho\Delta^2/2$, the excitation gap reading as $\Delta\simeq\Omega\sigma^{1/2}$ for $\sigma$ small.

Eq.(4) is very astonishing at first because it looks as the first two terms of the small $N$ expansion of the $N$-pair energy: Higher order $N$ terms seem missing! Actually, we have already reached these two first terms in a previous work \cite{PC} in which we analytically solved Richardson-Gaudin equations in the very dilute limit $N<<N_c$ where $N_c=\rho \varepsilon_c$ is the pair number over which single Cooper pairs would start to overlap. Since this number is far smaller than the BCS pair number $N_\Omega/2$, the procedure we have used to extract  the $R_j$'s from these equations is definitly not valid in the dense BCS limit. It moreover is so demanding that there were no hope to use it for higher order $N$ terms in order to at least check that they do cancel exactly.

 To prove that these higher order terms do not exist in the large volume limit, we have constructed a totally different procedure. We have found a way to reach the sum of $R_j$'s directly, without calculating these parameters separately, as we have done in our previous work. This is somehow necessary because the $R_j$'s are 2 by 2 complex conjugate - with one $R_j$ real for $N$ odd. When $N$ increases, these $R_j$'s run away from the real axis. This prevents their simple convergence for $N$ large.
By contrast, $\sum_j(R_j-E_1)/N\eta$ where 
\begin{equation}
\eta\equiv i\sqrt{N\gamma}=i\sqrt{2N/\rho\Omega\sigma}
\end{equation}
reduces to a degree-one $\eta$ polynomial due to a set of fundamental cancellations somewhat magic at first \cite{Tristan}.

Since the analytical resolution of Richardson-Gaudin equations for arbitrary $N$ and $V$ is a formidable math problem which stayed open for 50 years, some readers may wish to see the major steps of our procedure. These are given below, in a separate section, to be easily skipped by a more general audience ready to accept that we have indeed proved Eq.(4).

This very nice  solution for a model hamiltonian widely used in the literature for BCS superconductivity but also cold gases, is not only missing in all textbooks but it mostly brings some interesting new light in a field commonly considered as fully understood. Indeed, it provides a direct link between condensation energy and the Pauli exclusion principle, link out of reach when using the usual BCS ansatz in the grand canonical ensemble because $N$ is not a free parameter in this ensemble. By rewritting the condensation energy for $N$ pairs given in Eq.(5) as $E_{N}^{(0)}-E_N=N\varepsilon_c(N)$, we find that the average condensation energy per pair in the $N$-pair configuration, is proportional to the number of states available for pairing when $(N-1)$ pair states are already occupied. Indeed, for a total number of pairs $N_\Omega$ in this layer, the average condensation energy simply reads
\begin{equation}
\varepsilon_c(N)=\left[N_\Omega-(N-1)\right]\varepsilon^*.
\end{equation}
This result evidences that the unique consequence of increasing the number of pairs in the potential layer is to block more and more states in this layer, until all states are occupied: the correlation energy would then reduce to zero. In the Cooper problem, $N=1$ and $\varepsilon_c(1)=N_\Omega \varepsilon^*$ while in the usual BCS configuration $N=N_\Omega /2$ and the average condensation energy reduces to one half the single pair value.
Eq.(7) shows that Cooper pairs only "interact" through Pauli blocking. As a direct consequence, they can overlap without breaking, by contrast to excitons which dissociate into an electron-hole plasma through a Mott transition when overlap starts.

One important consequence of Eq.(7) is that the average pair energy cannot be identified, as commonly done in textbooks \cite{Fetter}, with the excitation gap much larger than the single pair value. The gap is the energy to break a pair. In the dense regime, this energy is far larger than just the energy of the broken pair due to Pauli induced many-body effects between the broken pair and the remaining $(N-1)$ unbroken pairs.

\textbf{Mathematical resolution of Richardson-Gaudin equations.
}
The exact energy of $N$ up and down spin electrons interacting via the BCS potential given in Eq.(1), reads \,\cite{Ric1,Gaud1,Zhu} as  $\sum_{j=1}^NR_j$ where the $R_j$'s are coupled through 
\begin{equation}
1=\sum_{\mathbf p}\frac{V w_{\mathbf p}}{2\varepsilon_{\mathbf p}-R_j}+\sum_{k\neq j}\frac{2V}{R_j-R_k}.
\end{equation}
We now list the major steps of our resolution.

$\bullet$ 
The first step is to note that for one pair, $E_1=R_1$, fulfills the above equation without the 
$k$ sum. By substracting this equation from Eq. (8), we get
\begin{equation}
0=\sum_{\mathbf p}w_{\mathbf p}\frac{R_j-E_1}{(2\varepsilon_{\mathbf p}-E_1)(2\varepsilon_{\mathbf p}-R_j)}+\sum_{k\neq j}\frac{2}{R_j-R_k}.
\end{equation}
The potential $V$ has disappeared from this equation: it is now hidden into $E_1$. 
We then set $R_j-E_1=\varepsilon_c z_j$ and expand the sum over $\mathbf p$ as a $z_j$ infinite series. By setting
\begin{equation}
\frac{2\varepsilon_c^{n}}{\rho(1{-}\sigma)}\sum_{\mathbf p}\frac{w_{\mathbf p}}{(2\varepsilon_{\mathbf p}-E_1)^{n+1}}=\frac{1{-}\sigma^{n}}{n(1{-}\sigma)}\equiv a_n,
\end{equation}
we find, for $\gamma$ defined in Eq.\,\eqref{e2}, that the $z_j$'s fulfill
\begin{equation}
0=\sum_{n=1}^\infty a_nz_j^{n}+\gamma \sum_{k\neq j}\frac{1}{z_j-z_k}.
\end{equation}
The unpleasant part of this equation is its second term.

$\bullet$ To get rid of it, we, in a second step, multiply the above equation by $z_j^\ell$ with $\ell=(0,1,2,...)$ and we sum over $j$. This gives  
\begin{equation}
0=\sum_{n=1}^\infty a_nZ_{n+\ell}+\gamma D_\ell,
\end{equation}
where the sums $Z_m$ and $D_m$ are defined as
\begin{equation}
Z_m=\sum_{j=1}^Nz_j^m,\qquad  D_m= \sum\sum_{j\neq k}\frac{z_j^m}{z_j-z_k}.
\end{equation}
The $N$-pair energy follows from $Z_1$.

Through the replacement of $z_j^m$ by $(z_j^m-z_k^m)/2$ in $D_m$, it is easy to show that $D_0=0$, $D_1=N(N\!-\!1)/2$, $D_2=(N\!-\!1)Z_1$, $D_3=(N\!-\!3/2)Z_2+Z_1^2/2$, and so on... As $Z_0=N$, the general expression of $D_m$ actually reads  
\begin{equation}
D_{m\geqslant1}= \frac{1}{2}\sum_{r=0}^{m-1} Z_rZ_{m-1-r}-\frac{m}{2}Z_{m-1}.
\end{equation}

 $\bullet$ The third step is to rescale $Z_m$ as $Z_m=N\eta^mX_m$ with $\eta$ defined in Eq.(6). Eqs.(12-14) then give the set of equations fulfilled by the $X_m$'s as 
 \begin{equation}
\sum_{n=1}^\infty a_n\eta^{n-1}X_{n+\ell}= \frac{1}{2}\sum_{r=0}^{\ell-1} X_rX_{\ell-1-r}-\frac{\ell}{2N}X_{\ell-1},
\end{equation}
 for $\ell\geqslant1$ while for $\ell=0$ the RHS reduces to zero. This readily shows that  
the $X_m$'s are $\eta$ series. Let us write them as $X_m=\sum_{q\geqslant0}x_{m,q}\eta^q$. Since $X_0=1$, we get $x_{0,0}=1$ and $x_{0,q\neq0}=0$. The other  $x_{m,q}$'s follow from identification of the $\eta^q$ terms in Eq.(15). A tedious but straightforward calculation shows that $X_1$ is an odd function of $\eta$ 
 \begin{equation}
X_1=-\eta\big(1{-}\frac{1}{N}\big)\frac{1{+}\sigma}{4}+\frac{1}{N}\,X_1^{\prime}.
\end{equation}
where $X_1^{\prime}$ depends on $(\eta, 1/N, \sigma)$ as
\begin{eqnarray}
X_1^{\prime}=\big(1{-}\frac{1}{N}\big)(1{+}\sigma)(1{-}\sigma)^2\Big\{y_{3,0}\eta^3+\big(y_{5,0}-
\frac{y_{5,1}}{N}\big)\eta^5
\nonumber\\
+\big(y_{7,0}-\frac{y_{7,1}}{N}+\frac{y_{7,2}}{N^2}\big)\eta^7+\cdots\Big\},\hspace{0.5cm}
\end{eqnarray}
  the $y_{m,k}$'s depending on $\sigma$ only.
This leads to
\begin{align}
E_N=\sum_{j=1}^NR_j=NE_1+\epsilon_cN\eta X_1 \hskip2.2cm \\\nonumber  
=NE_1+\frac{N(N-1)}{\rho}\frac{1+\sigma}{1-\sigma}+ \epsilon_c\eta X_1^{\prime}.\hskip1.0cm
\end{align}
Since the last term scales as $(\eta^4,\eta^6,\eta^8,\cdots)$, it gives volume free contributions to $E_N$ in $(N/\rho)^n$ with $n=(2,3,\cdots)$, in agreement with Eq.(4).
 
 $\bullet$ To be complete, the last important step of our resolution is to prove that \emph{all} corrections to the first two terms of $E_N$ are indeed volume free, i.e., $X_1$ reduces to $X_1^{(0)}=-\eta(1+\sigma)/4$ for large volume. To do it, we reconsider Eq.(15) without its last term
 \begin{equation}
 a_1X_{\ell+1}^{(0)}{+} a_2\eta X_{\ell+2}^{(0)}{+} a_3\eta^{2}X_{\ell+3}^{(0)}{+}\cdots= \frac{1}{2}\sum_{r=0}^{\ell-1} X_r^{(0)}X_{\ell-1-r}^{(0)}
\end{equation}
 The $X_{m}^{(0)}$'s, solution of this equation, are $\eta$ polynomials  
\begin{equation}
 X_{m}^{(0)}=x_{m,0}^{(0)} + x_{m,1}^{(0)}\eta +\cdots+x_{m,m}^{(0)}\eta^m.
\end{equation}
 the $x_{m,k}^{(0)}$ coefficients, defined for $0\leqslant k\leqslant m$ and nonzero for even $m+k$ only, being just the ones appearing in  
\begin{eqnarray}
 P_n(t,\sigma)=\frac{1}{2^n(n+1)!}\frac{d^n}{dt^n}\left[(t-\sigma)^n(t-1)^n \right]
 \\
 =x_{n,n}^{(0)} + x_{n+1,n-1}^{(0)}t +\cdots+x_{2n,0}^{(0)}t^n.
\end{eqnarray}
From $P_1(t,\sigma)=-(1+\sigma)/4+t/2$, we then find $X_1^{(0)}=x_{1,1}^{(0)}\eta=-\eta(1+\sigma)/4$ as we want.

The first key to check that the solution of Eq.(19) is given by Eq.(20) is to note that 
$P_n(t,\sigma)$'s is related to the Legendre polynomials $L_n(t)$ through 
 \begin{equation}
 P_n(t,\sigma)=\big(\frac{1-\sigma}{2}\big)^n\frac{1}{n+1}L_n\big(\frac{2t-1-\sigma}{1-\sigma}\big).
 \end{equation}
 The orthogonality of Legendre polynomials\,\cite{Gradstein} then gives for $0\leqslant k\leqslant n-1$ 
 \begin{equation}
 0=\frac{1}{1-\sigma}\int_\sigma^1t^kP_n(t,\sigma)dt.
 \end{equation}
By writting $P_n(t,\sigma)$ as in Eq.(22), it follows from this equation that $0=x_{n,n}^{(0)}a_{k+1}+x_{n+1,n-1}^{(0)}a_{k+2}+\cdots+x_{2n,0}^{(0)}a_{k+1+n}$, where $a_n$ is just the scalar appearing in Eq.(10). 
 
 The second key also follows from the link between $P_n(t,\sigma)$'s and the $L_n(t)$ Legendre polynomials. It is possible to show that 
 \begin{eqnarray}
 \frac{1}{1-\sigma}
  \int_\sigma^1\!\!\frac{P_n{(u,\sigma)}-P_n{(t,\sigma)}}{u{-}t}du \nonumber\\
  {=}\frac{1}{2}\sum_{r{=}0}^{n{-}1} P_r{(t,\sigma)} P_{n\!-\!1\!-\!r}{(t,\sigma)} \end{eqnarray}
the RHS reducing to zero for $n=0$. Eq.(22) inserted into the LHS of Eq.(25) gives this LHS as a $t$ polynomial $G_n(t)=x_{n+1,n-1}^{(0)}g_1(t)+\cdots+x_{2n,0}^{(0)}g_n(t)$ where $g_m(t)$ again depends on the $a_j$ scalar defined in Eq.(10) as $g_m(t)=\sum_{j=1}^ma_jt^ {m-j}$. For $n=1$, this in particular gives $G_1(t)=a_1x_{2,0}^{(0)}=1/2$ since $P_0(t,\sigma)=1$. 
 
 Using these two equalities, it becomes possible to show by identification that Eq.(20) fulfills Eq.(19). Details will be given in an extended version of this Letter.
   
   \textbf{Conclusion}: We here derive the energy of $N$ Cooper pairs by solving Richardson-Gaudin equations analytically for arbitrary $N$ and potential strength. We prove that, for large samples, the interaction part of the $N$-Cooper pair energy depends on $N$ as $N(N-1)$ only: higher order $N$ terms do not exist. As a result, the average Cooper pair binding energy $(E_N^{(0)}-E_N)/N$ \emph{linearly decreases with pair number from $N=1$ to the dense regime}, (see Eq.(5)), this energy being simply proportional to the number of empty states available for pairing in the potential layer. As a result, the pair binding energy cannot be identified with the gap, as commonly done.
   
   Our result fully supports the BCS result for the ground state energy obtained in the grand canonical ensemble when the potential layer extends symmetrically on both sides of the normal electron Fermi sea. One puzzling question still remains: Why the BCS ansatz leads to the \emph{exact} ground state energy for $N$ equal to half-filling since its projection onto the $N$-pair subspace corresponds to $\vert \psi _N^{(BCS)}\rangle {=}(B^\dagger)^N\vert F_0\rangle$ with \emph{all} pairs condensed into the \emph{same} state, while the exact Richardson-Gaudin ground state reads as $\vert \psi _{_{N}}\rangle= B^\dagger(R_1)\cdots B^\dagger(R_{_{N}})\vert F_0\rangle$ where
   \begin{align}\label{e7}
B^\dagger(R_j)=\sum_{\textbf{k}}\frac{w_\textbf{k}}{2\varepsilon_\textbf{k}-R_j}a^\dagger_{\textbf{k}\uparrow}a^\dagger_{-\textbf{k}\downarrow},
\end{align}
the $R_j$'s being  \emph{all different} due to Pauli blocking, as seen from the last term of Eq.(8)? Is this $N$-pair energy agreement also valid for correlation functions? We hope that this Letter will stimulate more works in connexion with the effects of the Pauli exclusion principle on paired electrons, in a field, conventional BCS superconductivity, commonly considered as fully understood.

We wish to thank Tony Leggett, Walter Pogosov, and Guojun Zhu for numerous discussions.

\end{document}